Andrei Nechayev

# About the Mechanism of Volcanic Eruptions

Geographic Department, Moscow State University

*A new approach to the volcanic eruption theory is proposed. It is based on a simple physical mechanism of the imbalance in the system "magma-crust-fluid". This mechanism helps to explain from unified positions the different types of volcanic eruptions. A criterion of imbalance and magma eruption is derived. Stratovolcano and caldera formation is analyzed. High explosive eruptions of the silicic magma is discussed*

Volcano is one of the most formidable and magnificent creations of nature. Volcanic eruptions sow everywhere fear, death and destruction. But people continue to live near volcanoes trying to understand and predict their behavior.

For centuries, the researchers are endeavoring to create a theory of volcanic activity not contradicting the most well-known facts of science. However, the volcanic eruptions that are available today for large-scale observations continue to put scientists in a deadlock. Clear and simple mechanism that can satisfactorily explain the diversity of volcanoes and types of their eruptions is still missing.

The mechanism of superheated "steam boiler" proposed at the beginning of the XIX century, was rejected because it does not find any theoretical or experimental evidence. Instead this idea a set of ideas concerning the excessive pressure of magma [1,2,3], its enrichment or degassing of volatile was proposed [1,3]. Nonlinear theory of catastrophes which is popular in recent years explained some important facts but was unable to answer the main questions[3]. The need for a simple universal mechanism of the eruption still not satisfied. It should be a mechanism capable to explain the slow, quiet outflow of lava and its gushing, emissions at supersonic speed of hot pyroclastic material, and the caldera forming catastrophes with cubic kilometers of erupted pumices and tuffs, Plinian eruptions with scorching clouds and perennial, hardly noticeable dome extrusions interrupted by catastrophic explosive paroxysms.

This paper describes a simple physical mechanism that allows from unified theoretical positions to explain different types of volcanic eruptions. Really this is the development and concrete definition of the old idea that water vapor (or some of its gas equivalent) can play a leading role in volcanic eruptions [4]. However, the specification of this idea became possible only after a new, previously unknown physical mechanism of imbalance between the hydrostatic liquid and the ideal gas was discovered and described. The essence of this mechanism, which can be called the "gas piston" mechanism, is that under certain conditions the contacting liquid and gas starts to uncontrolled relative motion: a column of liquid is erupting under the gas forcing. This mechanism being amenable to simple experimental verification was described in [6,7]. It explained not only the geyser phenomenon as such but many features of behavior of natural geysers of Kamchatka and Yellowstone. The "gas piston" mechanism is fundamental: the nature of liquid and gas as well as their density, viscosity and temperature do not play a principal role in it. Below we present a theoretical description of this mechanism in the simplified structure of "liquid - gas", we give the basic equations of dynamics and the instability criterion. And then we show how this mechanism may operate in the case of volcanic eruptions.

Let's assume that there is a vertical channel (rectangular or cylindrical) with solid walls which is filled to the brim with some liquid. There is also a volume filled with gas. Initially, liq-uid and gas are in equilibrium and have a direct contact area in accordance with Fig.1. Gas can leave its volume only through the contact with the liquid. Cross section of the channel is $S$, the volume of the gas chamber is $V$, the gas pressure is $p_g$. At a depth $z=0$ in the area of contact the hydrostatic pressure of liquid is equal to $p_0 + \rho g H$, where $\rho$ is the density of liquid, $p_0$ is the atmospheric pressure at $z=H$ (Fig.1). The gas pressure obeys the ideal gas law which in the case of an adiabatic process can be written:

$$p_g V^\gamma = A = const \qquad (1)$$

where $\gamma$ is the adiabatic coefficient for the gas ($\gamma = 1,4$ for water vapor). Since the change of gas volume is a process that runs much faster than all the processes of heat transfer, it can be considered as adiabatic.

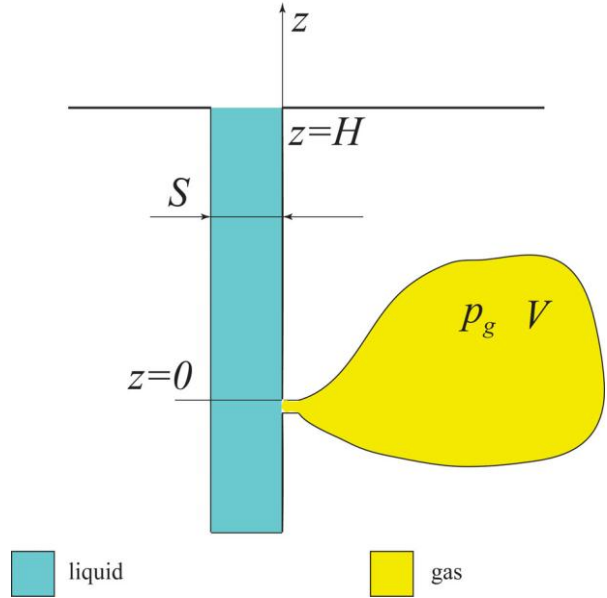

Fig.1 The scheme of "gas-liquid" structure.

Let's assume that the volume of a gas increases by a small amount $\Delta V$ due to the fact that some portion of the gas penetrates into the channel. In this case the same volume $\Delta V$ of liquid which is equal to $S \Delta z$ will be forced out from the channel where $\Delta z$ is the height reducing of the liquid column. The corresponding decrease of the hydrostatic pressure of liquid $\Delta p_l$ in the contact area is equal to:

$$\Delta p_l = -\rho g \Delta z = -\rho g \Delta V / S \qquad (2)$$

The pressure in the gas also decreases by an amount $\Delta p_g$ in accordance with equation (1):

$$\Delta p_g = \frac{\partial p_g}{\partial V} \Delta V = -\frac{\gamma A}{V^{\gamma+1}} \Delta V \qquad (3)$$

Thus, the more the volume of the gas chamber the less the pressure drop of gas during its expansion into the channel with the liquid. If the parameters of the structure $(H, S, V)$ are such that $|\Delta p_g| < |\Delta p_l|$ the hydrostatic pressure of the liquid in the contact area will decrease more rapidly than the pressure in the gas chamber and the gas will push the liquid from the channel like a piston. To define the critical parameters of the effect of "gas piston" we obtain the constant $A$ in (3) from the condition of equality of the gas pressure and the pressure of a liquid column in the contact area at the beginning of the process:

$$A/V^\gamma = p_0 + \rho g H \tag{4}$$

From (1), (3) and instability condition $|\Delta p_g| < |\Delta p_l|$ we obtain the criterion for the eruption of liquid from the channel:

$$V > \gamma S(H + p_0/\rho g) \tag{5}$$

For example, in the case of geyser eruption [7] liquid is a water, gas is a water vapor that accumulates during the water boiling in the underground cavity. Accordingly, $\gamma = 1,4$; $p_0/\rho g = 10m$.

The expression in the right part of (5) represents a critical volume $V^{cr}$. If the gas volume $V$ is much higher than $V^{cr}$, the pressure difference between gas and liquid will increase, the eruption of fluid will have an accelerated character until, for example, all liquid is ejected from the channel and the gas escapes. If $V < V^{cr}$, there is no instability, the gas penetrates into the liquid not pushing it and floats to the surface in the form of bubbles.

The same criterion of eruption can be obtained from the analysis of general conditions of the one-dimensional liquid flow in a channel in accordance with the principles of hydrodynamics. Indeed, the equation of motion in this case can be written as:

$$\frac{dv}{dt} = -\frac{1}{\rho}\frac{dp}{dz} - F_{fr}(v) \tag{6}$$

where $p_g$ is the liquid pressure in the section $z$, $F_{fr}(v)$ is the frictional force proportional to the average velocity in the case of laminar flow and to the square of the average velocity in the case of turbulent flow. For the pressure $p(z)$ we can write:

$$p = p_g - \rho g(H - z) - p_0 \tag{7}$$

where $p_g$ is the gas pressure, which is described by equation (1). Differentiating $p_g$ with respect to $z$ as an implicit function of the parameter $V$, we obtain:

$$\frac{dp_g}{dz} = \frac{\partial p_g}{\partial V}\frac{dV}{dz} = -\gamma \frac{A}{V^{\gamma+1}} S \tag{8}$$

as $\Delta V = S\Delta z$.

Substituting (7) and (8) in (6), we obtain the condition for the growth of velocity $v$ (the condition of the positive acceleration of the liquid) in the form:

$$\rho g(1 - \frac{\gamma S}{V}(H + p_0/\rho g)) - F_{fr}(v) > 0 \tag{9}$$

At $v = 0$ ($F_{fr}(0) = 0$) we get the criterion (5).

Two consequences principal for the volcanic eruptions can be derived from the analysis of equations (1) - (9) describing the mechanism of "gas piston". First, any dilatation of the channel located above the contact of gas and liquid does not change the criterion for instability (5) and is not an obstacle to the eruption. Second, the expansion of the gas and the corresponding reduction of its pressure can stop the eruption. Criterion for the eruption termination in contrast to the criterion (5) requires an analysis of the equations of motion (6), in which not only friction but also the pressure gradient is nonlinearly dependent on the velocity of the liquid.

Before passing to the description of "gas piston" mechanism in the case of volcanic eruptions, we discuss one of the major issues of volcanology: how deep magma penetrates the crust and rises to the surface, forming by way the so-called magma chambers.

There are a number of opinions on this issue [1]. Perhaps at some deep level the melt in the form of diapirs rise buoyantly as a less dense liquid. But the generally accepted cause of magma ascent is lithostatic pressure of thick solid crust on the underpinning it liquid magma [1,2]. Indeed, in certain areas of the crust (subduction zones, rift zones) magma along fractures and faults can rise to the surface of the earth. This ascent is provided by the overpressure of heavy rock on the melt with relatively less density. It is known that the density of magma varies weakly at depths greater than 1 km accounting for about 2.6 g/cm$^3$ (the most dense basaltic magma have about 2.8 g/cm$^3$). The density of the crust in the lithosphere increases from 2.0 g/cm$^3$ in the sedimentary layer (0 - 15 km) to 2.5 g/cm$^3$ in the granite (15 - 30 km) and 3.2 g/cm$^3$ in the lower layers of basalt (30 - 70 km) [2]. If the fracture in the lithospheric plate has an outlet to the surface, the magma will rise through it, until the pressure of the column balances the pressure of the surrounding solid rock. In other words:

$$\int_0^{H_0} \rho_c(z)dz = \int_0^{H_m} \rho_m dz \qquad (10)$$

where $\rho_m, \rho_c$ are the density of the magma and the crust, $H_0$ is the crust thickness, measured, for example, from the level of the asthenosphere, $H_m$ is the equilibrium height of the column of magma. The condition of lithostatic equilibrium would look like:

$$\bar{\rho}_c H_0 = \bar{\rho}_m H_m \qquad (11)$$

where $\bar{\rho}_m, \bar{\rho}_c$ are the corresponding averaged densities ($\bar{\rho}_c = \dfrac{1}{H_0}\int_0^{H_0} \rho_c dz$)

Depending on the density and thickness of crust layers the value of $H_m$ can be both larger and smaller than the value of $H_0$. In the first case the magma must overflow onto the surface of the earth. In the second case ($\bar{\rho}_c < \bar{\rho}_m$) magma can't reach the surface and stops at a certain depth. In the absence of accurate data on the densities of the crust at depths greater than 10 km, we can't be sure that the magma must be pushed to the surface if there is nothing to prevent. Perhaps this is just happening in the oceanic crust consisting mostly of dense basalt. In the continental crust all can be different: the thickness of sedimentary and granitic layers with density less than that of magma is great. It is likely that the magma can't reach the Earth's surface and stops at a certain level $H_m$.

It is believed that the magma chamber is formed at the depth where the magma has a neutral buoyancy ($\rho_m = \rho_c$) and supposedly has its magma column overpressure maximum [2]. However, it is logical to assume that the chamber can be formed where the pressure of the magma column exceeds strength $p_{st}$ of surrounding rocks. The condition of magma chamber formation at the level $H_{mc}$ would look like:

$$\rho_m g(H_m - H_{mc}) > p_{st} \qquad (12)$$

The tensile strength of granite rock is about 120 bar, that of sediment, such as limestone, is 90 bar, the corresponding strength of andesite is 60 bar only. Thus, the magma chamber could theoretically be formed at any depth, where the condition (12) is satisfied, for example, in porous and fractured areas on the borders of granite and sedimentary layers, of granite and basalt layers, etc. Being an indefinite time in the magma chamber without being able to flow to the surface and not getting fresh portions of the main mantle magma, the melt can change its physical and chemical properties, melting and dissolving into the surrounding rock and acquiring more and more silicic composition.

Volcanic eruptions that take place around the world demonstrate one thing: the magma leave their reservoirs and reaches the surface. Sometimes it happens at supersonic speed. How heavy viscous liquid gets such a big acceleration? Why it moves first slowly, first quickly? Why the breaks in the eruptions are sometimes short, sometimes long, sometimes periodic? To answer these questions one mechanism of magma ascent to the upper layers of the crust is not enough. Experts in the field of volcanology suggest that the leading role in the development of magma eruption the volatile components play: various gases dissolved in it under high pressure and released at the rise of magma to the surface [1,3]. These gases include, above all, water vapor, carbon dioxide, sulfur dioxide. Indeed, the solubility of volatiles in the magma increases sharply with increasing of pressure with pronounced jumps from 36% (at 30 kbar) to 20% (at 11 kbar) and then to 10% (at 10 kbar). The real quantity of volatiles dissolved in the magma at these depths, however, is unknown. Basaltic magma at the output contains about 1% of the volatile, silicic magma up to 6%. The rest may stand in the melt in the form of bubbles, which, according to scientists, creates the overpressure in the magma [1,3]. Dissolving and isolating of volatiles on the levels of fragmentation, as well as the processes of magma crystallization transform magma into a three-phase mixture which behavior analysis is rather complicated. Why the exsolving of volatiles can cause a great acceleration of the magma flow is not clear. The popular analogy with a bottle of Coca-Cola, in which the foamed mixture of drink and bubbles is thrown out after the opening is not evident. The acceleration of the liquid by the dissolved gas is keeping under pressure is possible only when you open the bottle suddenly with a sharp discharge of internal pressure and high magnitudes of pressure gradient. Really magma rises to the surface slowly, the pressure in it decreases gradually, bubbles stand out just as slowly. Apparently, the high rates of magma during eruption are the result of some other force action. Bubbles expansion in the magma itself is not able to accelerate the flow up to high, near-sonic speeds. They can blow up the magma and surrounding rocks if the pressure inside them becomes great when it is no way outside. But they can't accelerate magma and keep its speed for a long time! The acceleration of magma by its own bubbles is a sort of a well-known experience of Baron Munchausen, who pulled himself by the hair out of the swamp. Fragmentation can change the density and viscosity of the magma but not the pressure as volatiles existing as the Fluid are just there for a long time at supercritical temperature and pressure corresponding to the pressure inside the magma.

We assume that the "other force" providing a real acceleration of the magma and its eruption is caused by the direct contact with the area or layer containing Fluid, the so-called liquid, existing in the thick rocks under high pressure and high temperature. The role of Fluids in geology and mineralogy can't be overestimated. However, their participation in the geodynamic processes is underestimated.

The question of the Fluid genesis in the crust, especially at great depths, is highly debatable. We leave it to the side proceeding from the fact that areas containing Fluid really exist and you can find it, probably, at all depths. A convincing argument in favor of the Fluid may be the fact that in the subduction zones the oceanic crust which has the sedimentary layer saturated with water is pushing under the continental crust delivering water Fluid in the zone of active volcanism.

The simplest case of magma and Fluid contact is when magma conduit intersects the layer containing the Fluid (Fig.2). Perhaps this is a porous or fractured layer which has a relatively low strength and high permeability. If the magma with a temperature about 1300 K penetrates the Fluid area or approaches it there is a volume of Superheated Fluid which has not only a supercritical pressure but a supercritical temperature also. For water the critical magnitudes are respectively 220 bar and 647 K. The Superheated Fluid is compressible and obeys the ideal gas law (in a broad range of high temperature and pressure the difference between its behavior and the ideal gas behavior does not exceed 10% [8]). The aim of our work is to propose the Superheated Fluid to the role of an ideal gas and magma to the role of liquid in the above described "gas piston" mechanism.

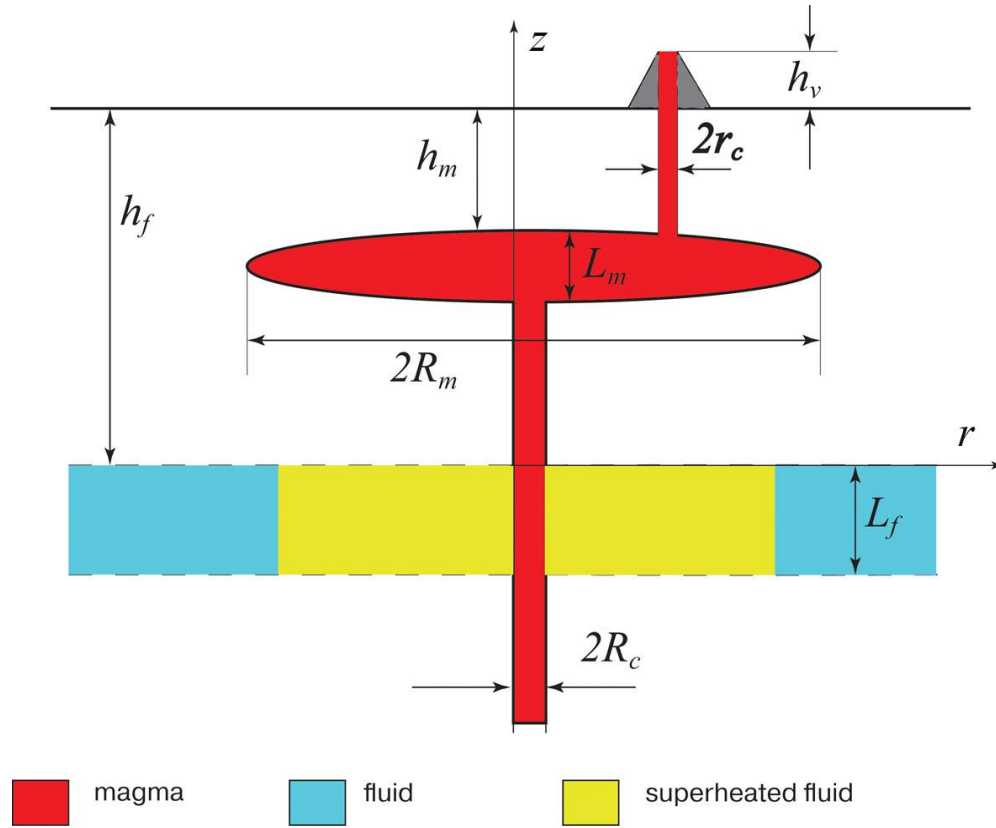

Fig.2 Scheme of magmatic system with Fluid layer

The area of Superheated Fluid besides its temperature $T_f$ and pressure $p_f$ (in the supercritical conditions the Fluid behaves as an ideal gas and its pressure and temperature are equalized in the volume due to convection) can be characterized by following parameters. This is the mass of Fluid $m_f$, the volume of Fluid $V_f$ (its density respectively $\rho_f = m_f / V_f$) as well as its molecular weight $M_f$. And finally its depth $h_f$ (level of bedding is $H_f$). Due to porosity and permeability of rocks containing Fluid its pressure is equal to the ideal gas pressure given in the form of Mendeleev-Clapeyron equation:

$$p_f V_f = \frac{m_f}{M_f} R T_f \qquad (13)$$

where $R$ is the Fluid universal gas constant.

The pressure of Fluid tends to a lithostatic pressure of the surrounding rocks. There are three options. First, the Fluid is compressed to this pressure (if the density of the Fluid is low). Second, the Fluid expands its volume and its pressure decreases to lithostatic one due to compression or destruction of surrounding crust areas. Third, there is the Fluid overpressure $\Delta p_f$ equal to:

$$\Delta p_f = p_f - \int_0^{h_f} \rho_c g dz \qquad (14)$$

If the Superheated Fluid comes into contact with magma conduit, the Fluid overpressure will push magma to the surface (by means of the dike or channel). In this case, the Fluid pressure will be equated with the pressure of the magma column tending to its maximum $\rho_m g h_f$. If the Fluid has sufficient volume it may cause the eruption of magma due to the "gas piston" mechanism. Let's consider the application of this mechanism of "magma-crust-fluid" imbalance for several types of volcanic eruptions.

In the area of contact the pressurized Fluid can push into the magmatic melt and float up in the form of bubbles because its density even at great depth is much less than the density of the magma (at a depth of 3 km at a pressure of 700 bar and a temperature of 1500 K, the density of water vapor is 100 kg/m3, that is 25 times less than the density of the magma [8]). Due to magma viscosity (especially silicic magma) the rise of Fluid bubbles will be rather slow: they will gather at the top of the magma column, bringing here its original pressure that is the actual pressure of the Fluid. Thus, the Superheated Fluid can form a kind of post magmatic gas "tip" which will push the rock like a spear and pave the magma way to the top. The same function can be implemented by the volatile fluids dissolved in the magma itself. The main mass of Superheated Fluid continues to prop up the magma column whose height above the Fluid approaches $h_f$.

Subsequent events depend on the ratio of the volume of the Superheated Fluid $V_f$ and conventional volume of the magmatic column $h_f S$, where $S$ is the column section at the place of contact with the Fluid layer (in the case of a cylindrical conduit $S = \pi R_c^2$). In accordance with the criterion (5) the movement of magma will go with the acceleration, if $V_f$ exceeds the critical volume $V_f^{cr}$, which is equal to:

$$V_f^{cr} = \gamma S(h_f + p_0/\rho_m g) \approx \gamma S h_f \qquad (15)$$

We emphasize that the value $p_0/\rho_m g$ for the magma is only 4 meters.

Thus, the condition of the eruption is as follows:

$$V_f > \gamma S h_f \qquad (16)$$

If (16) is not realized the magma will not accelerate. The Fluid bubbles will float upstairs in the magma column, will expand and push out some part of melt and burst near Earth's surface. However, this is not an eruption.

Assume now that the magma chamber is formed at a depth $h_m$ above the upper boundary of Fluid layer (Fig. 2). In this case the magma chamber will heat the rocks under it and after a certain time required for the heating of rock strata one has a vast area of Superheated Fluid with transverse dimensions corresponding to the horizontal size of the magma chamber. The thickness of this region will be determined by its temperature (supercritical water vapor temperatures must be higher than 647 K), but can't exceed the thickness of Fluid layer $L_f$. The zone of Superheat-ed Fluid will cover the magma conduit such as "collar" and its volume will be approximately equal $\pi L_f R_m^2$, since the chamber radius $R_m$ is much larger than the radius of the conduit $R_c$. If we take the conduit radius, the chamber radius and the depth of water Fluid layer respectively equal to 10 m, 2000 m and 5000 m, we find that the fulfilling of the criterion (16) requires Fluid layer thickness about 10 centimeters only!

The characteristic time of Fluid heating up to the supercritical temperature depends on the distance between the magma chamber and Fluid layer and on thermal diffusivity of rocks which is equal to $k/c\rho$ where $k$ is the thermal conductivity, $c$ is the specific heat, $\rho$ is the density. If we take typical values of these parameters for sedimentary rocks ($k = 2.5$ W / m K$^{-1}$, $c = 700$ J / kg K$^{-1}$, $\rho = 2500$ kg/m$^3$), the characteristic time $\tau_h$ of the temperature front moving from the chamber to the Fluid will be equal to approximately $\tau_h \approx 7 \cdot 10^5 (h_f - h_m)^2$ seconds. If the fluid is located 100 meters deeper than the magma chamber, $\tau_h$ will be 200 years about, if this distance becomes 1 km, $\tau_h$ will exceed 20,000 years. Recall that the "gas piston" mechanism is effective only when the volume of the Superheated Fluid exceeds the critical value and a magma column is reaching the surface.

We continue the description of the supposed eruption. The Superheated Fluid with $V_f > V_f^{cr}$ push magma to the surface by the help of "gas tip." Once a fracture was opened, the "gas cap" and the part of the magma column removed, the acceleration of magmatic mixture containing melt, fluid and gas begins (in this case the mixture is a liquid with a relatively low viscosity and Reynolds number exceeds its critical value). Velocity is stabilized by the turbulent friction of hot mixture on the walls of the conduit. The corresponding equation of hydrodynamics for an average speed of the mixture on the section of conduit between the fluid and the magma chamber would look like:

$$\frac{dv}{dt} = -\frac{1}{\rho_m}\frac{dp}{dz} - \lambda \frac{v^2}{R_c} \qquad (17)$$

where $\rho_m$ is the average density of the mixture (we suppose it equal to the density of the magma, although this is not true), $\lambda$ is the coefficient of turbulence, which depends in general on the Reynolds number. Taking into account that the pressure $p$ is equal to the difference between the pressure $p_f$ and the gravitational pressure of the melt column $\rho_m g z$, as well as the approach developed in the derivation of the formulas (7) - (9), we obtain:

$$\frac{dv}{dt} = -\frac{1}{\rho_m}\frac{\partial p_f}{\partial V}\frac{dV}{dz} + g - \lambda \frac{v^2}{R_c} \qquad (18)$$

where $V$ is the full volume of the Superheated Fluid, equal to the sum of the original volume $V_f$ and Fluid expansion volume in the magma conduit. In the section between the chamber and the Fluid layer ($0 < z < h_f - h_m - L_m$, Fig.2), where $V = V_f + Sz = V_f + \pi R_c^2 z$, taking into account equations (18) with (1) - (4) and the equality $p_f = \rho_m g h_f$ we obtain:

$$\frac{dv}{dt} = -\frac{\gamma g h_f S V_f^\gamma}{V^{\gamma+1}} + g - \lambda \frac{v^2}{R_c} \qquad (19)$$

From (19) follows that at the initial stage of magma motion through the conduit the acceleration of melt is independent of its density. The eruption begins when at $v = 0, z = 0$ the accelera-tion is positive. We get the same imbalance conditions (15), (16), where $S$ is the general section of the conduit (cylindrical or rectangular). Note that in case of dike with section $d \times w$, where $d$ is the width of the dike, usually not exceeding 1 m [1] and $w$ is its length which can reach several kilometers, we can have a failure of criterion (15), and for a given volume of Superheated Fluid the eruption through the dike will not happen. But through a cylindrical conduit with radius exceeding the width of the dike magma eruption can go with a significant acceleration.

The acceleration of the magma, as follows from (19), is proportional to $g(1-\gamma h_f S/V_f)$ and at $V_f \gg V_f^{cr}$ is close to the acceleration of gravity. The steady state magma acceleration decreases due to increasing of the turbulent friction on the conduit wall. Fixed rates of magma eruption is determined from (18) at $\partial v/\partial t = 0$:

$$g(1-\frac{\gamma h_f \pi R_c^2}{V_f}) = \lambda \frac{v^2}{R_c} \qquad (20)$$

As follows from (18) the main driving force of the eruption is a factor: $\partial p_f / \partial V = \gamma h_f V_f^{\gamma}/(V_f + V_{mf})^{\gamma+1}$, where $V_{mf}$ is the expansion volume of Superheated Fluid into the magmatic system, which includes a part of a conduit $(h_f - h_m - L_m)S$ and a volume, occupied by Fluid penetrating the magma chamber. At the eruption development the factor $\partial p_f / \partial V$ decreases, but at the initial stage of eruption (at $V_f \gg V_f^{cr}$) the Fluid pressure remains practically constant and equal to $\rho_m g h_f$. Apparently, the Superheated Fluid is introduced into the magma, accelerating it and throwing to the ground surface in the form of pyroclastic material. A cone of volcano begins to form from the cooled mixture of magma and gaseous fluid. This is a porous volcanic scoria which forms the primary cone of eruptions. The increase in total height of the magma column due to increase of the volcano cone is another factor that stabilizes the course of the eruption. If Fluid with $V_f \gg V_f^{cr}$ can reach the magma chamber, keeping its "native" pressure $\rho_m g h_f$, it resists the pressure of the magma column, equal to $\rho_m g(L_m + h_m + h_v)$ where $h_v$ is the height of the volcano cone, so the equation of motion of magma in the conduit above the magma chamber (Fig. 2) can be written as:

$$\frac{g(h_f - h_m - h_v)}{h_m + h_v} = \lambda \frac{v^2}{r_c} \qquad (21)$$

This equation implies that the eruption will weaken $(v \to 0)$ at $h_v \to (h_f - h_m)$, that is when the height of the cone of the volcano becomes close to the difference in the bedding depths of magma chamber and Fluid layer

Pressure reducing in the conduit due to Fluid expansion in the magma chamber or due to growth of the volcano cone causes the magma flow deceleration. The eruption turns to the effusive stage, the movement of magma becomes laminar, its rate is determined now by the Poiseuille equation: it is proportional to the square of the radius of the conduit and inversely proportional to the viscosity of the magma.

The condition for the termination of the eruption will be, probably, the equality of pressure of an enlarged (adiabatically) Fluid and the pressure of the magma column. It should be noted that the increasing of the Fluid volume $V_{mf}$ required for the termination of the eruption must be just equal to the corresponding volume of magma erupted from the magma chamber. This volume satisfies the condition:.

$$h_f(\frac{V_f}{V_f+V_{mf}})^\gamma = h_m + h_v \qquad (22)$$

If we assume that $h_v = 0$ (the cone is not growing, the material of the eruption dissipates into the atmosphere and spreads), we obtain the relation:

$$(V_f/(V_f+V_{mf}))^\gamma = h_m/h_f \qquad (23)$$

The closer the magma chamber to the Fluid layer ($h_f \to h_m$), the smaller the expansion of Fluid needed to stop the eruption, and, consequently, the eruption must end quickly. The more the distance separating the magma chamber and the Fluid layer, the more $V_{mf}$ and correspondingly longer the eruption time and more the volume of erupted rocks. If the volume of the magma chamber is comparable to or less than the volume of the Superheated Fluid, the Fluid pressure can completely devastate the chamber. This fact will tell us later the mechanism of Plinian and caldera forming eruptions. But now it is important that the eruption after it beginning has obvious physical reasons for its termination. The main cause among them is the pressure drop of the Fluid during its expansion.

In order that the eruption repeats and stratovolcano forms we need another important property of the Fluid. It is his ability to recover. Indeed, the concentration of the Fluid in the core of the eruption should be reduced or even felt to zero if all Fluid is ejected into the atmosphere. As a result, there must be a concentration gradient in the Fluid layer (the density gradient) and the corresponding diffusion fluxes of Fluid from the periphery to the magmatic conduit. There exists a characteristic time of recovery $\tau_f$ of the Fluid density in the zone of overheating, devastated by the eruption. Maybe this time creates the real interval between eruptions. Frequency of eruptions, therefore, may be determined by the time of filling the magmatic system of fresh magma, the diffusion time of the Fluid from the periphery and the time of its heating to extreme temperatures (perhaps a new warm-up is not required, as the cooling of rocks after the eruption is a more long process than the feeding of the system with fresh magma [9]). The necessary critical mass of Fluid will be achieved due to diffusion of the Fluid from the cool periphery and its entering into the zone of overheating. Contact of Fluid with a magma chamber can occur at $V_f \gg V_f^{cr}$ already that immediately provokes a new eruption which goes as above described scenario: the release of gas and pyroclastic material, cone growth, reducing of the Fluid pressure, transition to the expiration of volcanic lava. If the pressure of the magma column inside the scoria cone exceeds the tensile strength of rocks composing it, there is a lateral breakout of lava, which is typical for areal eruptions [2,5]. To resume the eruption from the summit crater, the pressure of the Fluid should be no less than $\rho_m g(h_f + h_v)$ to raise the magma to the crater and carry out the primary drop of magma column pressure. If the Fluid pressure remained stable, the eruption may continue by the parasitic crater on the surface (through new fracture). If the Fluid pressure is increasing with no lateral breakthrough the volcano will continue to grow. The magnitude of this growth during given eruption can't exceed the value of $(h_f - h_m)$. Thus, the maximum height of the volcano and its activity should be determined by a set of parameters of the Fluid $\rho_f, h_f, \tau_f, V_f$. The deeper the Fluid layer, the more its density, the more its distance from the magma chamber, the more powerful eruption could be.

For the eruption beginning it is necessary to satisfy criterion (15) and to provide the possibility of small pressure drop of magma column, which may occur during the breakthrough of magma through the slope of the volcano, In this case, the eruption can take place simultaneously from the central and parasitic craters.

Obviously, the activity of volcano should weaken when the exhaustion of the Fluid layer comes. The driving force of the eruption disappears, the volcano, fall "asleep." The magma in the magma chamber is not updated with fresh magma from the mantle and acquires silicic character. Cooling and crystallization of magma in the chamber can last for hundreds of thousands of years. But everything changes when at a great depth under the magma chamber a new Fluid layer is activated. If it is deeper by 1 km than the chamber, his warming-up to supercritical temperatures will require more than 20 000 years. The additional 1 km requires more than 80,000 years. But when the Fluid layer receives the desired temperature and the critical volume, it may cause a powerful eruption, which can devastate a huge magma chamber, filled with silicic magma and cause the formation of the caldera. The power of the eruption, as the dynamic equation (18) gives, depends on the depth of the Fluid layer as compared with the depth of the magma chamber. Owing to the large amount of Superheated Fluid its pressure remains virtually unchanged during the expansion in the magma conduit and the Fluid can maintain its extremely high pressure during penetration in the magma chamber and its devastation. Mixed with the Fluid magma is ejected from the magma chamber in the form of cubic kilometers of pumice and tuff. After the failure of the upper layer of the crust and the formation of the caldera the structure of the magmatic system may be compromised. The new chamber can be formed at new location and new depth. But the activity of the volcano is already restored because the deep Fluid layer is activated, or more precisely its relatively small part, located under the magma chamber. Restoration of of the Fluid concentration can take place simultaneously with the resumption of the activities of the volcano. In place of the caldera a new stratovolcano may begin to grow. An activation of two or more Fluid layers (one is "small", others are "deep") can take place simultaneously, combining the formation of stratovolcanoes and calderas.

If the volume of the Superheated Fluid is large and substantially exceed the critical volume (15), if the bedding depth of the Fluid layer is large too, the Fluid pressure may completely devastate the magmatic system. The Superheated Fluid is pulled out like a genie from a bottle. Its speed is supersonic, the temperature drops from 1500 K due to its expansion but no more than one or two hundred degrees. Flying through the channel, he tears off from the walls tiny particles of melt and turns into a deadly scorching cloud. The volume of fluid can be enormously large, as being in an extremely compressed state in deep layer, he goes after the eruption in the free atmosphere (remember the eruption of Mount Pinatubo and 30 million tons of sulfur dioxide). Obviously, the output of the Fluid is the catastrophic finale of Plinian eruption, the final, which was preceded by a long eruption of pyroclastics and devastation of magma chamber.

The theory which claims to universality must explain the most complex and controversial aspects of the phenomenon. Let's try to interpret with the help of our model the high-explosive eruptions of the volcanoes. In the history of volcanology some of them are well-known under the names of Bezymianny volcano (Kamchatka, 1956), Shiveluch volcano (Kamchatka, 1964), Mount St Helens (USA, 1980) [1,5] These devastating eruptions usually are preceded by a slight strengthening of the volcano: enhanced fumarole activity (Bezymianny), extrusive dome growth (Shiveluch), the deformation of the slopes (St Helens). The eruption, lasting usually a few hours, accompanied by a shock wave, a huge cloud of ash (up to 45 km altitude), pyroclastic and mud flows. We believe that this type of eruption is the result of interaction between Fluid layers and the magmatic system in the later stages of the history of the stratovolcano, when its magma becomes silicic. Suppose that the "responsibility" for the volcano "birth" is carried by the Fluid layer located at depth $h_f$ (Fig. 3) with a volume $V_f$ greater than the critical volume $\gamma S h_f$. This Fluid layer is not exhausted, but eruptions became rare, since for their beginning it is not enough overpressure of restored Fluid. Perhaps eruptions were affected by the height of the volcano close to its maximum. Magma that is stored in chambers and conduits acquired silicic composition, a high viscosity and the ability to solidify with a slight cooling. This led to the formation at the top of the conduit a sort of magmatic "plug", i.e. a part of the magmatic column of height $h$ with an extremely viscous magma (Fig. 3).

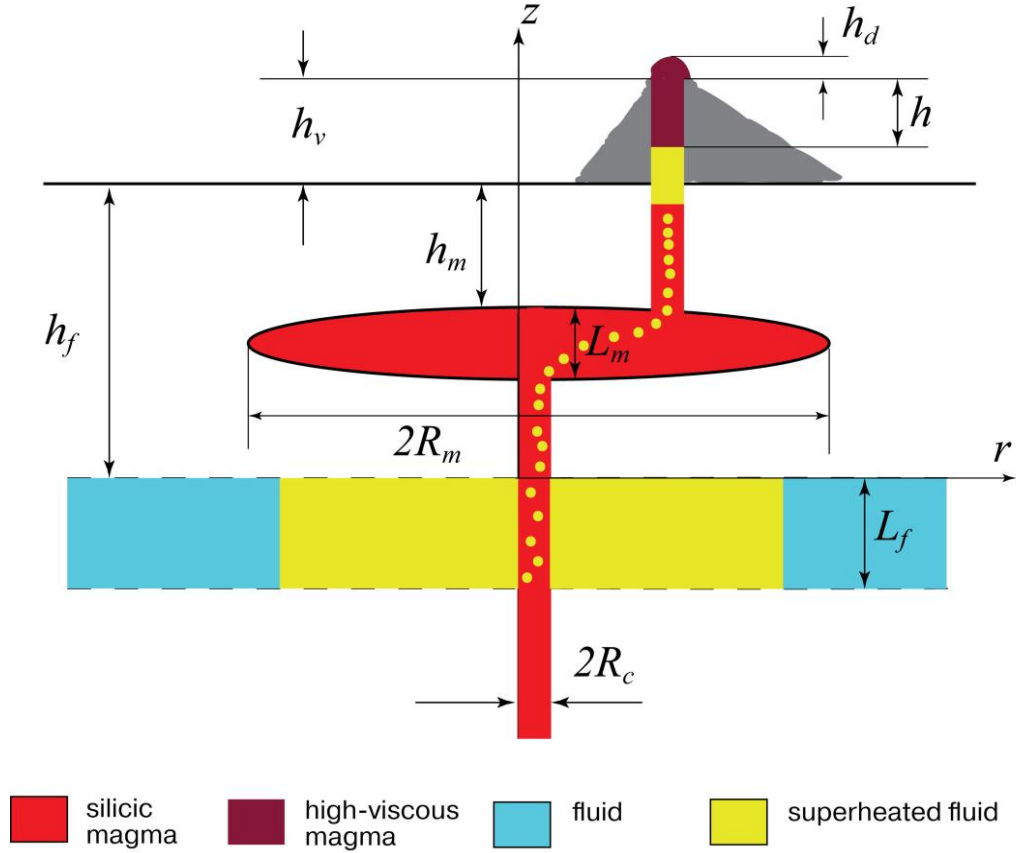

Fig.3 Scheme of magmatic system with Fluid layer for silicic magma and explosive eruptions

In accordance with the above described "gas piston" mechanism, the slow outflow of magma occurs as a result of the instability (at $V_f > V_f^{cr}$), and fixed rate should be determined by an equation similar to (20) for laminar flow:

$$g(1 - \frac{\gamma h_f \pi R_c^2}{V_f}) = \frac{\mu v}{8 R_c^2} \qquad (24)$$

Since the kinematic viscosity of the silicic magma is large, the rate of flow of magma is very small, but the eruption is going: It is expressed in the growth of extrusive domes and rock falls on their slopes. Expiration of the magma occurs under Fluid pressure, which in this case is equal to $\rho_m g(h_f + h_v + h_d)$ where $h_v$ is the height of the active cone volcano and $h_d$ is the height of extrusive dome (Fig. 3). In the contact zone the Fluid is pressed into the magma conduit, and the Fluid bubbles rise to the top because its density is ten times less than the density of the magma. The ascent of bubbles is also very slow due to the viscosity of the magma. Inside the "plug" the motion of bubbles is not possible, they should build up under it the sort of "gas tip" in the form of a large bubble, which occupies the entire cross section of the conduit and has almost the same pressure as its "native" Fluid at its depth. The bubble keeps the pressure of the Fluid, because the "plug" does not allow him to expand and to rise to the top of magma column as it occurs during eruptions of basaltic magma. In accordance with the formula (21) and since the eruption (extrusion) of the magma began in the volcano already formed, it must be stabilized when the height of the extrusive dome reaches its maximum possible value equal to $(h_f - h_m)$. Note that the growth of extrusive domes in active craters of volcanoes Bezymianny and Shiveluch has stopped at 500 m height [5].

Thus, the "gas tip" having the pressure of the deep Fluid moves through the conduit along with the magma approaching the surface. Through cracks in the magmatic column and fissures surrounding the channel the Fluid can go into the atmosphere and decrease its pressure. However, if the outflow of the Fluid will be equal to its inflow in the magma conduit at the depth of the Fluid layer, the pressure in the "gas tip" will be retained. Thus, the high pressure of deep strata is transported through the Fluid bubbles up to the crater of the volcano and stopped up by the viscous magma.

In order to magmatic structure preserves its integrity, the pressure in the "gas tip" shall not exceed the sum of the magma "plug" weight and a total strength of the channel walls adhesion with surrounding rocks. Both forces are proportional to "plug" height $h$ that either decreases with magma extrusion, or remain unchanged if the crystallization and solidification of magma during its cooling at the surface will be ahead of the extrusion process. The probability that the tensile strength of the structure will be once exceeded is quite large. At some point, "gas tip" will receive the contact with the atmosphere and there will be a blast. Its power depends on the pressure $\rho_m g h_f$ of the Fluid. It may be intense outflow of Superheated Fluid to form pyroclastic flows and ash clouds. Similar paroxysms occur 1-2 times over several years in active extrusive domes of Shiveluch and Bezymianny [

The process of explosive eruption preparing depends on the depth of Fluid layer ($h_f$ gives the pressure of the Fluid) and on the depth at which a "gas tip" formed (i.e. on the height $h$ of the "plug"). If the latter was below ground level, the critical, destructive, Fluid pressure $p_f^{cr}$ will be determined only by the vertical direction (the direction of the magma channel):

$$p_f^{cr} = \rho_m g(h+h_d) + p_{st} \frac{2\pi R_c h}{\pi R_c^2} = \rho_m g(h+h_d) + p_{st} \frac{2h}{R_c} \qquad (25)$$

where $p_{st}$ is the ultimate strength of adhesion of the channel walls with the surrounding rocks. At the magma extrusion (firstly, with the growth of extrusive dome and then after its reaching the maximum height with possible following collapse) the critical pressure (25) can change according to changes in magnitude $(h+h_d)$. If the Fluid pressure exceeds a critical one (25), there will be a blast. As a result of "plug" ejection the crater is forming with the diameter depending on the depth at which the imbalance of forces occurred.

If in the process of magma extrusion the top of the "gas tip" penetrates the cone of the volcano (above the ground level), the risk of explosion in the lateral direction increases, as the pressure of the Fluid is not opposed to the weight of the magma "plug" but to the thickness of rocks separating the conduit from the slope of the volcano. The critical pressure for this lateral explosion may be less than the vertical one, especially for the andesitic rocks, which tensile strength does not exceed 60 bar.

After the explosion of the "gas tip" and removing of the "plug" the extrusion of viscous magma through the conduit can resume, as Fluid pressure at depth $h_f$ remains. A new extrusive dome will begin to grow which is typical for the explosive eruptions.

Some more about one interesting type of explosive eruptions, which can be explained using the aforementioned theoretical arguments. Suppose that under magma chamber there are two Fluid layers at different depths. Of course, the "deep" one can be activated much later than the "shallow" layer located closer to the magma chamber. Consequently, when two Fluid layers begin to function together, there may be two cycles of eruptions that intersect with each other. The first cycle is determined by the "shallow" Fluid layer and it has one characteristic time recovery of the Fluid, a second, the "deep" Fluid layer, will determine more powerful eruptions with longer intervals, as its recovery time should be substantially greater. Cycles will overlap. This type of eruption may be observed on the volcano Shiveluch and Bezymianny [4,5]. Explosive eruptions of ash clouds with pyroclastic flows occur at them every few years and, apparently, are inside the more powerful and disaster cycle, which have the period of 100-200 years for Shiveluch, and about 1000 years for Bezymianny.

Thus, the combination of magma chambers with differently bedded Fluid layers through the "gas piston" mechanism can provide a large variety of theoretical options of volcanic activity.

# Conclusions

In this paper a simple physical mechanism that can serve as a basis for explaining the different types of volcanic eruptions is proposed. It is assumed that the acceleration of magma and its eruption is caused by imbalance in the contact zone of magmatic conduit and Fluid layer at depths exceeding 1 km. The Superheated Fluid (may be water vapor Fluid) at the pressure greater than 220 bar and temperatures above 647 K obeys the ideal gas law. According to this law the more the primary Fluid volume the less the decrease of Fluid pressure during its expansion into magma conduit. If the Superheated Fluid penetrates into magma system and displaces some small part of magma to the surface, and its volume at the same time exceeds a certain critical magnitude, there is a pressure difference at the contact zone, and Fluid begins to push the magma out of the conduit like a piston. The analysis showed that the value of critical volume is approximately equal to $\gamma S h_f$, where $\gamma$ is the adiabatic factor of the Fluid (for water vapor $\gamma = 1,4$), $S$ is the cross section of the magma conduit, $h_f$ is the bedding depth of the Fluid layer. During volcanic eruption the Fluid does work and expands, its pressure and density decrease, the eruption ends. To repeat it and to form a stratovolcano the system needs some time to restore the critical volume of Fluid in the zone of overheating, devastated by the eruption. This may occur due to diffusion of Fluid from the periphery. Perhaps this time corresponds to the interval between the eruptions. After fluid layer depletion the activity of the volcano should be interrupted. But if under magma chamber at big depth there exists another Fluid layer, it can achieve supercritical temperature tens thousands years after the formation of a stratovolcano and cause the powerful eruption that can devastate the magma chamber resulting in the caldera formation. After the eruption of magma and pyroclastic material the Superheated Fluid can escape. Flying through the channel at the supersonic speed, it could carry away small particles of melt, turning into the scorching ash cloud. The eruption of the fluid is a Plinian eruption finale, preceded by the formation of caldera and stratovolcano. All these events result from the excess of the critical conditions and primary imbalance in the "magma-crust-fluid" system.

References

1. Pinkerton H., Wilson L., Macdonald R. The transport and eruption of magma from volcanoes: a review. *Contemporary Physics.* 2002, v.43, n.3, p. 197-210
2. Fedotov S.A. Magmatic feeding systems and mechanism of volcano eruptions. Nauka, Moscow, 2006, 456 p. (in Russian)
3. Melnik O., Sparks R.S.J. Nonlinear dynamics of lava dome extrusion. *Nature.* v.402,4 November 1999, p.37-41
4. Modern and Holocene volcanism in Russia. (ed. Laverov N.S.). Nauka, Moscow, 2005, 608 p. (in Russian)
5. Nechayev A. Kamchatka. Volcanoes Kingdom. LOGATA, Moscow, 2008, 200p.
6. Nechayev A. New physical mechanism of Geyser operating: theory and its confirmation based on many years observations in the Valley of Geysers in Kamchatka. – IAVCEI 2008 General Assembly, Reykjavik, Iceland 17-22 August 2008, Abstracts, p.97.
7. Nechayev A. About the mechanism of geyser eruption. arXiv:1204.1560v1, 2012, 13p.
8. Zubarev V.N., Kozlov A.D., Kuznetsov V.M. Thermophysical properties of technically important gases at high temperatures and pressures. Energoatomizdat, Moscow, 1989, 232 p.
9.Polyansky O.P., Reverdatto V.V. The role of fluid flow during the heat and mass transfer induced by basaltic intrusions in sedimentary basins. *In the book:* Fluids and Geodynamics, Nauka, Moscow, 2006, p.219-243